\documentclass[prl,twocolumn]{revtex4}
\usepackage{amsmath}
\usepackage{amssymb}
\usepackage{float}
\usepackage{bm,times}
\usepackage{graphicx}
\newcommand{\be}{\begin{equation}}
\newcommand{\ee}{\end{equation}}
\newcommand{\bea}{\begin{eqnarray}}
\newcommand{\eea}{\end{eqnarray}}

\begin{document}

\title{Understanding the superfluid phase diagram in trapped Fermi gases}

\date{\today}

\author{Qijin  Chen$^1$, C. A. Regal$^2$, M. Greiner$^2$, D. S. Jin$^2$,
and K. Levin$^1$}

\affiliation{$^1$ James Franck Institute and Department of Physics,
  University of Chicago, Chicago, Illinois 60637}

\affiliation{$^2$ JILA, National Institute of Standards and
Technology Quantum Physics Division and University of Colorado and
Department of Physics, University of Colorado, Boulder, Colorado
80309-0440}

\begin{abstract}

Trapped ultracold Fermi gases provide a system that can be tuned
between the BCS and BEC regimes by means of a magnetic-field
Feshbach resonance.  Condensation of fermionic atom pairs in a
$^{40}$K gas was demonstrated experimentally by a sweep technique
that pairwise projects fermionic atoms onto molecules.  In this
paper we examine previous data obtained with this technique that
probed the phase boundary in the temperature-magnetic field plane.
Comparison of the $^{40}$K data to a theoretically computed phase
diagram demonstrates good agreement between the two.
\hfill \textbf{\textsf{cond-mat/0512596}}
\end{abstract}


\maketitle

%----------------------------------------------------

The field of ultracold Fermi gases has seen enormous progress in
the years since the first observation of Fermi degeneracy
\cite{Jin}.  Upon variation of a magnetic field $B$, Feshbach
resonances provide a means of controlling the strength of
interactions between fermionic atoms, which is characterized by
the $s$-wave scattering length $a$. The nature of the resultant
superfluidity is expected to vary continuously \cite{Leggett} from
BEC ($a >0$) to BCS ($a< 0$) as a function of magnetic-field
detuning from the resonance. In the BEC regime, evidence for
condensation was obtained \cite{Jin3,Grimm2,Ketterle2} by
observing a bimodal distribution of the momentum profile, a
standard technique originally developed for bosonic gases.

On the BCS side of resonance the situation is experimentally more
complicated.  To demonstrate condensation a momentum projection
technique based on fast sweeps into the BEC regime was introduced
\cite{Jin4}.  Detailed time-dependent studies suggest that the
sweeps used are sufficiently rapid that a condensate cannot be
created during this sweep process.  The presence of a condensate
\textit{after} a sweep then provides strong support for the
existence of a condensate \textit{before} the sweep on the BCS
side of resonance.  In this way the first Fermi gas
normal-superfluid (NS) phase diagram was obtained experimentally
for $^{40}$K \cite{Jin4} and later for $^6$Li \cite{Ketterle3}.
Additional experiments in $^6$Li have since added to the evidence
for fermionic superfluidity, including, collective mode
observations \cite{Thomas2,Grimm3}, thermodynamic measurements
near unitarity \cite{ThermoScience} and, most conclusively, the
demonstration of quantized vortices \cite{KetterleV}. These data
in conjunction with the sweep experiments serve to further
constrain the NS boundary.

The purpose of this paper is to present a comparison of this
important phase boundary measured in fast sweep experiments
involving $^{40}$K to a theoretical computation of the NS
boundary.  Recent theoretical work examining the entropy of the
trapped gas in the BCS-BEC crossover now makes it possible to make
this comparison \cite{ChenThermo}. Thus we present previous data
\cite{Jin4} in a new way and show that the experimentally obtained
condensate fraction in $^{40}$K provides a good measure of this
phase boundary.  While the emphasis of Ref. \cite{Jin4} was on
providing evidence for the condensation of atom pairs in an
ultracold Fermi gas, here we show that these data moreover provide
a universal NS phase diagram, as a function of temperature and
interaction strength, that can be quantitatively compared with
theory.

In making this quantitative comparison several important issues
need to be considered. First we need to consider the fast sweep
technique and what information it can provide.  The essential and
unique feature of this technique is that it can provide direct
information about the condensate fraction by means of a
measurement of a bimodal distribution in the particle density or
momentum profile. For the fermionic regime this information is
obtained \cite{Jin4,Ketterle3} by rapidly sweeping the magnetic
field beginning on the BCS side of resonance to the BEC, or $ a >
0$, side of resonance where time-of-flight imaging can be used to
measure the momentum distribution of the weakly bound molecules.
The projecting magnetic field sweep is completed on a time scale
that allows molecule formation but is still too brief to allow
additional pairs to condense.

In the $^{40}$K experiment it was observed that the fast sweep
resulted in significant number loss \cite{Jin3,Jin4}, presumably
because of the relatively short lifetime of the molecules away
from resonance \cite{Jin7}. The measured condensate fraction,
which is defined as the number of condensed molecules divided by
the total number of molecules observed after the fast sweep, could
be affected by this loss. However, the loss process is almost
certainly density dependent and thus one expects only suppression
(and never enhancement) of the condensate fraction.  Furthermore
one expects a large loss only for large condensate fractions.
Therefore, the NS phase line (separation between zero condensate
fraction and a finite value) obtained in the experiment should be
relatively unaffected. When plotted as a function of the
temperature, it corresponds to the threshold curve below which a
finite fraction of the molecules is observed to have near zero
momentum.

Second, we note that a general difficulty in these Fermi gas
experiments is the lack of model-independent thermometry in the
strongly interacting regime. Therefore experiments typically rely
on the combination of temperature measurements made away from
resonance and slow adiabatic sweeps to the strongly interacting
regime. In this paper we use a superscript ``$0$" to denote
quantities measured away from the Feshbach resonance in the weakly
interacting Fermi gas regime.  The temperature relative to the
Fermi temperature $(T/T_F)^0$ is determined from surface fits to
absorption images of the gas taken after expansion from the trap.
We have checked \cite{Regal2} that this yields accurate
temperature measurements down to $(T/T_F)^0\approx0.1$. Below this
temperature measuring $(T/T_F)^0$ this way becomes more difficult
due to the fact that the momentum distribution of the Fermi gas
approaches the $T=0$ limit. The NS phase diagram also depends on
the adiabaticity of the magnetic-field sweep toward resonance.
Studies of the condensate fraction as a function of sweep rate
\cite{Jin3} suggest that the sweep toward resonance is
sufficiently slow. More recent studies involving double ramps to
the resonance and then back away suggest that extra heating during
the ramp is not significant on the BCS side of resonance
\cite{Jin6}.

Third, for comparison with theory the magnetic-field values should
be converted to the dimensionless parameter $1/k_Fa$, which
reflects the strength of the pairing interaction in BCS-BEC
crossover theories. Here $k_F$ is the Fermi wavevector at the trap
center, and $a$ is the two-body $s$-wave scattering length between
fermionic atoms. For $a$ we use a previous measurement of the
scattering length as a function of magnetic field \cite{Jin4}, and
for the Fermi wavevector we use $k_F$ measured in the weakly
interacting regime, $k_F^0$.  Here $E_F^0 \equiv k_BT_F^0=
\hbar^2(k_F^0)^2/2m$, where $E_F^0$ is the noninteracting Fermi
energy.  As defined here,
$k_F^0=(2m\overline{\omega}/\hbar)^{1/2}(3N_a)^{1/6}$, where $N_a$
is the total number of atoms and
$\overline{\omega}=(\omega_x\omega_y\omega_z)^{1/3}$ is the
geometric mean angular trap frequency.  Results for the NS phase
diagram are then plotted in terms of $(T/T_F)^0$ and the
dimensionless parameter $1/k_F^0a$. The phase diagram plotted this
way is nearly universal and should be applicable to $^6$Li as
well. However, because of the large resonance width $^6$Li
experiments may have some difficulty in reaching the
noninteracting Fermi gas regime, which sets the temperature scale
we use here.

Our theoretical calculations are based on the finite temperature
formalism described in Ref.~\cite{ChenThermo}.  Comparisons with
the experimental data require not only an understanding of the
BCS-BEC crossover system \cite{ourreview,ReviewJLTP} at finite
$T$, including the behavior of $T_c$ in a trap and the
temperature-dependent superfluid density $N_s$, but also an
understanding of the entropy $S$.  Note that $(T/T_F)^0$ is
essentially a measure of the entropy of the gas.  We take as a
ground state the usual BCS-Leggett wave function and include
excitations of finite momentum pairs. At unitarity, the resulting
thermodynamics has been shown to compare favorably to experiment
in Ref. \cite{ThermoScience}. More generally, the excitations that
contribute to the temperature dependence of the energy can be
attributed to both ``bosonic'' and fermionic degrees of freedom,
which are strongly inter-dependent. The latter are associated with
a finite excitation gap, which in turn is a measure of the
presence of fermion pairs or ``bosons''. Similarly, the superfluid
density reflects both the bosonic and fermionic contributions
found in the thermodynamics. The entropy is dominated by fermionic
excitations in the BCS regime and bosonic excitations in the BEC
regime. In an adiabatic sweep from BCS to BEC, the temperature
increases.  A detailed theory of the thermometry for adiabatic
sweeps is presented in Ref.~\cite{ChenThermo}.

\begin{figure}
\includegraphics[width=3.2in,clip]{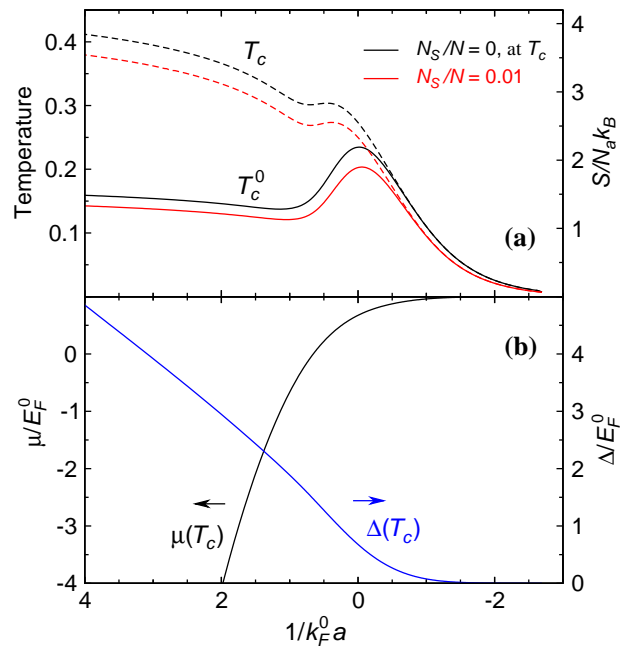}
\caption{(color online) (a) Physical temperature $T/T_F^0$ (dashed
  curves), effective temperature $(T/T_F)^0$, and entropy per
  particle $S/N_ak_B$ (solid curves) at the superfluid transition (black
  curves) and $N_s/N=0.01$ (red curves). (b) $\mu(T_c)$ (black
  curve) and $\Delta(T_c)$ (blue curve) at the trap center as
  functions of $1/k_F^0a$. In (a), the solid lines represent both
  $(T/T_F)^0$ and $S/N_ak_B \propto T^0$, where $N_a$ is the total
  number of atoms of both spins.  $(T/T_F)^0$ is the temperature
  measured in the noninteracting Fermi gas limit.} \label{fig:Tc}
\end{figure}

This approach leads to a self-consistent set of equations for the
fermionic atoms and Cooper pairs throughout the crossover. The
transition temperature $T_c$ is associated with a vanishing of the
chemical potential of the pairs.  The magnitudes of this
transition temperature (in a trap) are similar to those found
elsewhere \cite{Strinati4} using a different ground state, where
less is currently known about the superfluid density and
thermodynamics.

We now address the implications for the phase diagram.
In the experiment the temperature of the gas is determined in the
noninteracting Fermi gas limit at high field; we denote this
temperature by $T^0$.  The destination field is accessed
adiabatically by a slow magnetic-field sweep. Using the theory in
Ref.~\cite{ChenThermo}, we calculate the entropy at different
magnetic fields and temperatures. In this way, we can associate
the physical temperature $T$ with the effective temperature $T^0$.
We then calculate the condensate fraction $N_0/N$, here identified
with the superfluid density $N_s/N$, as a function of temperature
$T$ or $T^0$ and of magnetic field $B$.  The latter parameter is
appropriately characterized by the dimensionless variable
$1/k_F^0a$, which provides a measure of the strength of pairing
interaction.  At a given field this parameter varies with the
Fermi temperature.

In Fig.~\ref{fig:Tc}(a), we show the results for the superfluid
transition temperature $T_c$ (black dashed line) and its
corresponding value $T_c^0$ (black solid line)for an isentropic
sweep into the Fermi gas regime as functions of $1/k_F^0a$.  In a
similar fashion, we plot the physical (red dashed line) and
effective temperatures (red solid line) corresponding to
$N_s/N=0.01$.
In Fig.~\ref{fig:Tc}(b), we plot the fermionic chemical potential
$\mu(T_c)$ and the excitation gap $\Delta(T_c)$ at the trap center
as a function of $1/k_F^0a$. When the chemical potential is
negative the system can be viewed as ``bosonic", whereas when
$\mu$ is positive it is ``fermionic."

\begin{figure}
  \includegraphics[width=2.8in,clip]{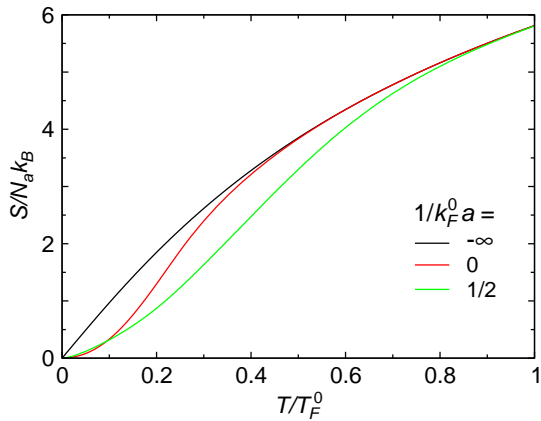}
  \label{fig:entropy} \caption{(color online) Entropy per particle
    $S/N_ak_B$ as a function of physical temperature $T/T_F^0$ for
    $1/k_F^0a= -\infty$ (black line), $0$ (red line), and $1/2$ (green
    line), representing the ideal Fermi gas, unitary, and strongly interacting BEC cases,
    respectively.}
\end{figure}

Because the entropy for a noninteracting gas at low $T/T_F^0$ is
nearly linearly dependent on the temperature, one can conclude
that $T_c^0$ is essentially proportional to the entropy at the
transition $S(T_c)$. The latter is labeled on the right hand axis
of Fig.~\ref{fig:Tc}(a). It follows that as a natural consequence
of an isentropic sweep, $T_c^0$ is reduced substantially from the
physical $T_c$ except in the BCS regime. As can be seen from the
figure, this reduction is dramatic in the BEC regime ($1/k_F^0a >
0.7$) and persists essentially to unitarity ($1/k_F^0 a = 0$). One
can understand this reduction as reflecting the presence of
bosonic degrees of freedom at $T_c$. Once noncondensed bosons or
preformed pairs are present at the temperature of their
condensation, the entropy curve for $S(T)$ for $T \leq T_c$ drops
substantially below its counterpart for a noninteracting Fermi gas
at the same temperature. One can alternatively say that when $T_c$
and $T_c^0$ are significantly different a normal state excitation
gap or ``pseudogap" \cite{ourreview,ReviewJLTP,Jin5} is present at
$T_c$.  In the fermionic regime ($\mu > 0$) and at the transition
temperature, this pseudogap is parametrized by $\Delta(T_c)$,
which is also shown in Fig.~\ref{fig:Tc}(b) and should be viewed
as an alternative measure of bosonic degrees of freedom.  One can
see that the difference between $T_c$ and $T_c^0$ reflects rather
nicely the behavior of $\Delta(T_c)$ as a function of $1/k_F^0a$.
Beginning at unitarity and moving towards the BEC regime,
$\Delta(T_c)$ increases rapidly, reflecting the rapid increase in
the bosonic degrees of freedom. This leads then to a strong
reduction from $T_c$ to $T_c^0$ and explains the existence of the
maximum seen in $T_c^0$.

\begin{figure}[tb]
\includegraphics[width=3.4in,clip]{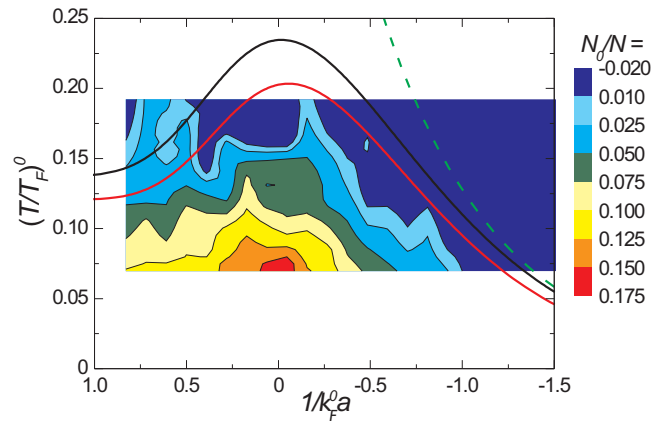}
\caption{(color) Phase diagram of $^{40}$K. A contour plot of the
  measured condensate fraction $N_0/N$ as a function of $1/k_F^0a$ and
  effective temperature $(T/T_F)^0$ is compared with theoretically
  calculated contour lines at $N_s/N=0$ (at the superfluid transition,
  black curve) and 0.01 (red curve).  The experimental data have an
  overall systematic uncertainty of approximately 0.1 in $1/k_F^0a$.
  The overall trend of the experimental contour of $N_0/N=0.01$ and the
  theoretical line for $N_s/N=0.01$ are in good agreement.  The dashed
  line represents the naive BCS result $T_c/T_F^0 \approx 0.615
  e^{\pi/2k_F^0a}$.  Here all temperatures are measured in the Fermi gas
  regime.}
\label{fig:Phase}
\end{figure}

To illustrate these effects, in Fig. 2 the temperature dependence
of the entropy $S(T)$ is shown for selected values of $1/k_F^0a$
representing the Fermi gas, unitary, and BEC cases. Here we see
that the noninteracting gas result at low $T/T_F^0$ is close to a
straight line, and that as the system becomes more strongly
interacting, $S(T)$ deviates from this line with decreasing
temperature. This deviation sets in once the temperature goes
below the pair formation temperature, $T^*$.

We are now in a position to compare our calculated phase diagram
with experimental measurements.  In Fig.~\ref{fig:Phase}, we
replot the measured phase diagram from Ref.~\cite{Jin4} as a
function of $1/k_F^0a$ and overlay our theoretical curves.  The
top (black) curve corresponds to the theoretical calculation for
$(T_c/T_F)^0$, whereas the remaining (red) line is the effective
temperature $(T/T_F)^0$ corresponding to the superfluid fraction
$N_s/N= 0.01$.  We present both theoretical curves because (as can
be seen from the disproportionate breadth of the contour swath for
$0<N_s/N<0.01$ in Fig. 4, see also Ref.~\cite{ChenThermo}), the
superfluid density has a flat tail close to the transition
temperature due to trap inhomogeneity effects. Consequently
experimental noise may add a large uncertainty to the temperature
for which $N_s/N= 0$; the $1\%$ contour should be an
experimentally more robust boundary.

In general, the phase boundary for $N_s/N=0.01$ is in good
agreement with the experimentally measured phase boundary for
$N_0/N =0.01$. However, in the near-BEC regime there are a few
experimental data points that show a finite condensate fraction
above the theoretical transition line. The two well-known
weaknesses of the mean-field approach may be partly responsible
for this discrepancy: Both the overestimate of the interboson
scattering length in the BEC regime and underestimate of the value
of the "beta" factor at unitarity lead to a slightly
underestimated peak density of the trap profile, which in turn
leads to an underestimate of $T_c$. Also, in the experiment, the
slow sweeps that extend to the BEC side of the resonance are not
perfectly adiabatic. Moreover, the experiment finds less than
100\% conversion of atoms to molecules or pairs in this regime
\cite{Hodby}.

\begin{figure}[tb]
\includegraphics[width=3.4in,clip]{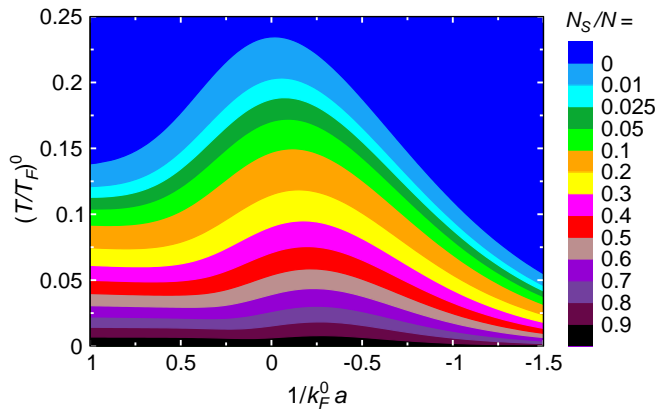}
\caption{(color) The theoretically computed equilibrium phase diagram
  and contour plot of the superfluid density $N_s/N$ as a function of
  $(T/T_F)^0$ and $1/k_F^0a$.  } \label{fig:PhaseTh}
\end{figure}

In Fig.~\ref{fig:PhaseTh} we plot the theoretical phase diagram and
contour plot for $N_s/N$, which should be appropriate to a variety of
different experiments.  We can compare this calculation with its
experimental counterpart in Fig.~\ref{fig:Phase}.  When comparing the
theoretical values of $N_s/N$ at the lowest temperature [$(T/T_F)^0
\approx 0.07$] accessed experimentally, we find that, at unitarity, the
theoretical value is about two times as large as in the $^{40}$K
experiment.  This difference may be attributed to a number of factors.
First, the condensate fraction may be partly destroyed during the
projection sweep process.  A possible sweep-induced reduction of the
condensate fraction has been addressed elsewhere \cite{Perali2,Ho3}.  In
Refs.~\cite{Perali2,Ho3}, however, no attempt was made to distinguish
between the actual physical temperature and that measured in the weakly
interacting regime.  Second, sweeps may not be 100\% adiabatic; minor
heating is known to be present for sweeps from the noninteracting Fermi
gas to the BEC regime.  Finally, in the fermionic and unitary regimes
the condensate fraction is not a unique parameter and varies with the
particular observation under consideration.  In the present
measurements, it is not known precisely how well the superfluid density
represents this fraction.  This uncertainty does not affect the NS phase
boundary.  The superfluid density, however,  should be regarded as an upper
bound to the condensate fraction. As an example, a recent Monte Carlo
calculation defined a $T=0$ condensate fraction that is nearly a factor of
two lower than $N_s/N$ at unitarity \cite{Giorgini}. Given these
factors, the agreement between theory and experiment is reasonably good.

In summary, we have shown that previous measurements of
Ref.~\cite{Jin4} of the normal state-superfluid phase boundary in
$^{40}$K are in good agreement with theoretical calculations.  In
addition to reinforcing the theoretical approach taken here, this work
adds support to previous claims that fast sweep experiments do, indeed,
provide a reliable indication of the phase boundary.  A feature of the
predicted phase boundary is that, when it is plotted in terms of the
temperature $T^0_c$ in the noninteracting regime, there is a maximum
near unitarity as a function of $1/k_F^0a$.  Indications for this
maximum have also been observed \cite{Ketterle3} in $^6$Li. Here one
finds that the biggest condensate fraction occurs near unitarity, as is
predicted in Fig.~\ref{fig:PhaseTh}.

%-
We thank Eric Cornell for providing the impetus for this paper and
Cheng Chin for very helpful comments.  This work was supported by
NSF, NASA, and NSF-MRSEC Grant No.~DMR-0213745. C.A.R.
acknowledges support from the Hertz foundation.

\vspace*{-3ex}
\bibliographystyle{apsrev}
%\bibliography{Review3}

\begin{thebibliography}{23}
\vspace*{-3ex}

\expandafter\ifx\csname natexlab\endcsname\relax\def\natexlab#1{#1}\fi
\expandafter\ifx\csname bibnamefont\endcsname\relax
  \def\bibnamefont#1{#1}\fi
\expandafter\ifx\csname bibfnamefont\endcsname\relax
  \def\bibfnamefont#1{#1}\fi
\expandafter\ifx\csname citenamefont\endcsname\relax
  \def\citenamefont#1{#1}\fi
\expandafter\ifx\csname url\endcsname\relax
  \def\url#1{\texttt{#1}}\fi
\expandafter\ifx\csname urlprefix\endcsname\relax\def\urlprefix{URL }\fi
\providecommand{\bibinfo}[2]{#2}
\providecommand{\eprint}[2][]{\url{#2}}

\bibitem[{\citenamefont{DeMarco and Jin}(1999)}]{Jin}
\bibinfo{author}{\bibfnamefont{B.}~\bibnamefont{DeMarco}} \bibnamefont{and}
  \bibinfo{author}{\bibfnamefont{D.~S.} \bibnamefont{Jin}},
  \bibinfo{journal}{Science} \textbf{\bibinfo{volume}{285}},
  \bibinfo{pages}{1703} (\bibinfo{year}{1999}).

\bibitem[{\citenamefont{Leggett}(1980)}]{Leggett}
\bibinfo{author}{\bibfnamefont{A.~J.} \bibnamefont{Leggett}}, in
  \emph{\bibinfo{booktitle}{Modern Trends in the Theory of Condensed Matter}}
  (\bibinfo{publisher}{Springer-Verlag}, \bibinfo{address}{Berlin},
  \bibinfo{year}{1980}), pp. \bibinfo{pages}{13--27}.

\bibitem[{\citenamefont{Greiner et~al.}(2003)\citenamefont{Greiner, Regal, and
  Jin}}]{Jin3}
\bibinfo{author}{\bibfnamefont{M.}~\bibnamefont{Greiner}},
  \bibinfo{author}{\bibfnamefont{C.~A.} \bibnamefont{Regal}}, \bibnamefont{and}
  \bibinfo{author}{\bibfnamefont{D.~S.} \bibnamefont{Jin}},
  \bibinfo{journal}{Nature} \textbf{\bibinfo{volume}{426}},
  \bibinfo{pages}{537} (\bibinfo{year}{2003}).

\bibitem[{\citenamefont{Bartenstein
  et~al.}(2004{\natexlab{a}})\citenamefont{Bartenstein, Altmeyer, Riedl,
  Jochim, Chin, Hecke, and R}}]{Grimm2}
\bibinfo{author}{\bibfnamefont{M.}~\bibnamefont{Bartenstein}},
  \bibinfo{author}{\bibfnamefont{A.}~\bibnamefont{Altmeyer}},
  \bibinfo{author}{\bibfnamefont{S.}~\bibnamefont{Riedl}},
  \bibinfo{author}{\bibfnamefont{S.}~\bibnamefont{Jochim}},
  \bibinfo{author}{\bibfnamefont{C.}~\bibnamefont{Chin}},
  \bibinfo{author}{\bibfnamefont{D.}~\bibnamefont{Hecke}}, \bibnamefont{and}
  \bibinfo{author}{\bibfnamefont{G.}~\bibnamefont{R}}, \bibinfo{journal}{Phys.
  Rev. Lett.} \textbf{\bibinfo{volume}{92}}, \bibinfo{pages}{120401}
  (\bibinfo{year}{2004}{\natexlab{a}}).

\bibitem[{\citenamefont{Zwierlein et~al.}(2003)}]{Ketterle2}
\bibinfo{author}{\bibfnamefont{M.~W.} \bibnamefont{Zwierlein}}
  \bibnamefont{et~al.}, \bibinfo{journal}{Phys. Rev. Lett.}
  \textbf{\bibinfo{volume}{91}}, \bibinfo{pages}{250401}
  (\bibinfo{year}{2003}).

\bibitem[{\citenamefont{Regal et~al.}(2004{\natexlab{a}})\citenamefont{Regal,
  Greiner, and Jin}}]{Jin4}
\bibinfo{author}{\bibfnamefont{C.~A.} \bibnamefont{Regal}},
  \bibinfo{author}{\bibfnamefont{M.}~\bibnamefont{Greiner}}, \bibnamefont{and}
  \bibinfo{author}{\bibfnamefont{D.~S.} \bibnamefont{Jin}},
  \bibinfo{journal}{Phys. Rev. Lett.} \textbf{\bibinfo{volume}{92}},
  \bibinfo{pages}{040403} (\bibinfo{year}{2004}{\natexlab{a}}).

\bibitem[{\citenamefont{Zwierlein et~al.}(2004)\citenamefont{Zwierlein, Stan,
  Schunck, Raupach, Kerman, and Ketterle}}]{Ketterle3}
\bibinfo{author}{\bibfnamefont{M.~W.} \bibnamefont{Zwierlein}},
  \bibinfo{author}{\bibfnamefont{C.~A.} \bibnamefont{Stan}},
  \bibinfo{author}{\bibfnamefont{C.~H.} \bibnamefont{Schunck}},
  \bibinfo{author}{\bibfnamefont{S.~M.~F.} \bibnamefont{Raupach}},
  \bibinfo{author}{\bibfnamefont{A.~J.} \bibnamefont{Kerman}},
  \bibnamefont{and} \bibinfo{author}{\bibfnamefont{W.}~\bibnamefont{Ketterle}},
  \bibinfo{journal}{Phys. Rev. Lett.} \textbf{\bibinfo{volume}{92}},
  \bibinfo{pages}{120403} (\bibinfo{year}{2004}).

\bibitem[{\citenamefont{Kinast et~al.}(2004)\citenamefont{Kinast, Hemmer, Gehm,
  Turlapov, and Thomas}}]{Thomas2}
\bibinfo{author}{\bibfnamefont{J.}~\bibnamefont{Kinast}},
  \bibinfo{author}{\bibfnamefont{S.~L.} \bibnamefont{Hemmer}},
  \bibinfo{author}{\bibfnamefont{M.~E.} \bibnamefont{Gehm}},
  \bibinfo{author}{\bibfnamefont{A.}~\bibnamefont{Turlapov}}, \bibnamefont{and}
  \bibinfo{author}{\bibfnamefont{J.~E.} \bibnamefont{Thomas}},
  \bibinfo{journal}{Phys. Rev. Lett.} \textbf{\bibinfo{volume}{92}},
  \bibinfo{pages}{150402} (\bibinfo{year}{2004}).

\bibitem[{\citenamefont{Bartenstein
  et~al.}(2004{\natexlab{b}})\citenamefont{Bartenstein, Altmeyer, Riedl,
  Jochim, Chin, Denschlag, and Grimm}}]{Grimm3}
\bibinfo{author}{\bibfnamefont{M.}~\bibnamefont{Bartenstein}},
  \bibinfo{author}{\bibfnamefont{A.}~\bibnamefont{Altmeyer}},
  \bibinfo{author}{\bibfnamefont{S.}~\bibnamefont{Riedl}},
  \bibinfo{author}{\bibfnamefont{S.}~\bibnamefont{Jochim}},
  \bibinfo{author}{\bibfnamefont{C.}~\bibnamefont{Chin}},
  \bibinfo{author}{\bibfnamefont{J.}~\bibnamefont{Denschlag}},
  \bibnamefont{and} \bibinfo{author}{\bibfnamefont{R.}~\bibnamefont{Grimm}},
  \bibinfo{journal}{Phys. Rev. Lett.} \textbf{\bibinfo{volume}{92}},
  \bibinfo{pages}{203201} (\bibinfo{year}{2004}{\natexlab{b}}).

\bibitem[{\citenamefont{Kinast et~al.}(2005)\citenamefont{Kinast, Turlapov,
  Thomas, Chen, Stajic, and Levin}}]{ThermoScience}
\bibinfo{author}{\bibfnamefont{J.}~\bibnamefont{Kinast}},
  \bibinfo{author}{\bibfnamefont{A.}~\bibnamefont{Turlapov}},
  \bibinfo{author}{\bibfnamefont{J.~E.} \bibnamefont{Thomas}},
  \bibinfo{author}{\bibfnamefont{Q.~J.} \bibnamefont{Chen}},
  \bibinfo{author}{\bibfnamefont{J.}~\bibnamefont{Stajic}}, \bibnamefont{and}
  \bibinfo{author}{\bibfnamefont{K.}~\bibnamefont{Levin}},
  \bibinfo{journal}{Science} \textbf{\bibinfo{volume}{307}},
  \bibinfo{pages}{1296} (\bibinfo{year}{2005}), \bibinfo{note}{published online
  27 January 2005; doi:10.1126/science.1109220}.

\bibitem[{\citenamefont{Zwierlein et~al.}(2005)\citenamefont{Zwierlein,
  Abo-Shaeer, Schirotzek, and Ketterle}}]{KetterleV}
\bibinfo{author}{\bibfnamefont{M.~W.} \bibnamefont{Zwierlein}},
  \bibinfo{author}{\bibfnamefont{J.~R.} \bibnamefont{Abo-Shaeer}},
  \bibinfo{author}{\bibfnamefont{A.}~\bibnamefont{Schirotzek}},
  \bibnamefont{and} \bibinfo{author}{\bibfnamefont{W.}~\bibnamefont{Ketterle}},
  \bibinfo{journal}{Nature} \textbf{\bibinfo{volume}{435}},
  \bibinfo{pages}{170404} (\bibinfo{year}{2005}).

\bibitem[{\citenamefont{Chen et~al.}(2004)\citenamefont{Chen, Stajic, and
  Levin}}]{ChenThermo}
\bibinfo{author}{\bibfnamefont{Q.~J.} \bibnamefont{Chen}},
  \bibinfo{author}{\bibfnamefont{J.}~\bibnamefont{Stajic}}, \bibnamefont{and}
  \bibinfo{author}{\bibfnamefont{K.}~\bibnamefont{Levin}}
  (\bibinfo{year}{2004}), \bibinfo{note}{arXiv:cond-mat/0411090; \prl
  \textbf{95}, 31 DEC 2005.}

\bibitem[{\citenamefont{Regal et~al.}(2004{\natexlab{b}})\citenamefont{Regal,
  Greiner, and Jin}}]{Jin7}
\bibinfo{author}{\bibfnamefont{C.~A.} \bibnamefont{Regal}},
  \bibinfo{author}{\bibfnamefont{M.}~\bibnamefont{Greiner}}, \bibnamefont{and}
  \bibinfo{author}{\bibfnamefont{D.~S.} \bibnamefont{Jin}},
  \bibinfo{journal}{Phys. Rev. Lett.} \textbf{\bibinfo{volume}{92}},
  \bibinfo{pages}{083201} (\bibinfo{year}{2004}{\natexlab{b}}).

\bibitem[{\citenamefont{Regal and Jin}(2005)}]{Regal2}
\bibinfo{author}{\bibfnamefont{C.~A.} \bibnamefont{Regal}} \bibnamefont{and}
  \bibinfo{author}{\bibfnamefont{D.~S.} \bibnamefont{Jin}}
  (\bibinfo{year}{2005}), \bibinfo{note}{unpublished}.

\bibitem[{\citenamefont{Regal et~al.}(2005)\citenamefont{Regal, Greiner,
  Giorgini, Holland, and Jin}}]{Jin6}
\bibinfo{author}{\bibfnamefont{C.~A.} \bibnamefont{Regal}},
  \bibinfo{author}{\bibfnamefont{M.}~\bibnamefont{Greiner}},
  \bibinfo{author}{\bibfnamefont{S.}~\bibnamefont{Giorgini}},
  \bibinfo{author}{\bibfnamefont{M.}~\bibnamefont{Holland}}, \bibnamefont{and}
  \bibinfo{author}{\bibfnamefont{D.~S.} \bibnamefont{Jin}},
  \bibinfo{journal}{Phys. Rev. Lett.} \textbf{\bibinfo{volume}{95}},
  \bibinfo{pages}{250404} (\bibinfo{year}{2005}).

\bibitem[{\citenamefont{Chen et~al.}(2005{\natexlab{a}})\citenamefont{Chen,
  Stajic, Tan, and Levin}}]{ourreview}
\bibinfo{author}{\bibfnamefont{Q.~J.} \bibnamefont{Chen}},
  \bibinfo{author}{\bibfnamefont{J.}~\bibnamefont{Stajic}},
  \bibinfo{author}{\bibfnamefont{S.~N.} \bibnamefont{Tan}}, \bibnamefont{and}
  \bibinfo{author}{\bibfnamefont{K.}~\bibnamefont{Levin}},
  \bibinfo{journal}{Phys. Rep.} \textbf{\bibinfo{volume}{412}},
  \bibinfo{pages}{1} (\bibinfo{year}{2005}{\natexlab{a}}).

\bibitem[{\citenamefont{Chen et~al.}(2005{\natexlab{b}})\citenamefont{Chen,
  Stajic, and Levin}}]{ReviewJLTP}
\bibinfo{author}{\bibfnamefont{Q.~J.} \bibnamefont{Chen}},
  \bibinfo{author}{\bibfnamefont{J.}~\bibnamefont{Stajic}}, \bibnamefont{and}
  \bibinfo{author}{\bibfnamefont{K.}~\bibnamefont{Levin}}
  (\bibinfo{year}{2005}{\natexlab{b}}), \bibinfo{note}{arXiv:cond-mat/0508603}.

\bibitem[{\citenamefont{Perali et~al.}(2004)\citenamefont{Perali, Pieri,
  Pisani, and Strinati}}]{Strinati4}
\bibinfo{author}{\bibfnamefont{A.}~\bibnamefont{Perali}},
  \bibinfo{author}{\bibfnamefont{P.}~\bibnamefont{Pieri}},
  \bibinfo{author}{\bibfnamefont{L.}~\bibnamefont{Pisani}}, \bibnamefont{and}
  \bibinfo{author}{\bibfnamefont{G.~C.} \bibnamefont{Strinati}},
  \bibinfo{journal}{Phys. Rev. Lett.} \textbf{\bibinfo{volume}{92}},
  \bibinfo{pages}{220404} (\bibinfo{year}{2004}).

\bibitem[{\citenamefont{Greiner et~al.}(2005)\citenamefont{Greiner, Regal, and
  Jin}}]{Jin5}
\bibinfo{author}{\bibfnamefont{M.}~\bibnamefont{Greiner}},
  \bibinfo{author}{\bibfnamefont{C.~A.} \bibnamefont{Regal}}, \bibnamefont{and}
  \bibinfo{author}{\bibfnamefont{D.~S.} \bibnamefont{Jin}},
  \bibinfo{journal}{Phys. Rev. Lett.} \textbf{\bibinfo{volume}{94}},
  \bibinfo{pages}{070403} (\bibinfo{year}{2005}).

\bibitem[{\citenamefont{Hodby et~al.}(2004)\citenamefont{Hodby, Thompson,
  Regal, Greiner, Wilson, Jin, Cornell, and Wieman}}]{Hodby}
\bibinfo{author}{\bibfnamefont{E.}~\bibnamefont{Hodby}},
  \bibinfo{author}{\bibfnamefont{S.~T.} \bibnamefont{Thompson}},
  \bibinfo{author}{\bibfnamefont{C.~A.} \bibnamefont{Regal}},
  \bibinfo{author}{\bibfnamefont{M.}~\bibnamefont{Greiner}},
  \bibinfo{author}{\bibfnamefont{A.~C.} \bibnamefont{Wilson}},
  \bibinfo{author}{\bibfnamefont{D.~S.} \bibnamefont{Jin}},
  \bibinfo{author}{\bibfnamefont{E.~A.} \bibnamefont{Cornell}},
  \bibnamefont{and} \bibinfo{author}{\bibfnamefont{C.~E.}
  \bibnamefont{Wieman}}, \bibinfo{journal}{Phys. Rev. Lett.}
  \textbf{\bibinfo{volume}{94}}, \bibinfo{pages}{120402}
  (\bibinfo{year}{2004}).

\bibitem[{\citenamefont{Perali et~al.}(2005)\citenamefont{Perali, Pieri, and
  Strinati}}]{Perali2}
\bibinfo{author}{\bibfnamefont{A.}~\bibnamefont{Perali}},
  \bibinfo{author}{\bibfnamefont{P.}~\bibnamefont{Pieri}}, \bibnamefont{and}
  \bibinfo{author}{\bibfnamefont{G.~C.} \bibnamefont{Strinati}},
  \bibinfo{journal}{Phys. Rev. Lett} \textbf{\bibinfo{volume}{95}},
  \bibinfo{pages}{010407} (\bibinfo{year}{2005}).

\bibitem[{\citenamefont{Diener and Ho}(2004)}]{Ho3}
\bibinfo{author}{\bibfnamefont{R.}~\bibnamefont{Diener}} \bibnamefont{and}
  \bibinfo{author}{\bibfnamefont{T.-L.} \bibnamefont{Ho}}
  (\bibinfo{year}{2004}), \bibinfo{note}{arXiv:cond-mat/0404517}.

\bibitem[{\citenamefont{Astrakharchik et~al.}(2005)\citenamefont{Astrakharchik,
  Boronat, Casulleras, and Giorgini}}]{Giorgini}
\bibinfo{author}{\bibfnamefont{G.~E.} \bibnamefont{Astrakharchik}},
  \bibinfo{author}{\bibfnamefont{J.}~\bibnamefont{Boronat}},
  \bibinfo{author}{\bibfnamefont{J.}~\bibnamefont{Casulleras}},
  \bibnamefont{and} \bibinfo{author}{\bibfnamefont{S.}~\bibnamefont{Giorgini}},
  \bibinfo{journal}{Phys. Rev. Lett.} \textbf{\bibinfo{volume}{95}},
  \bibinfo{pages}{230405} (\bibinfo{year}{2005}).

\end{thebibliography}

\end{document}